\documentclass[%
reprint,
superscriptaddress,
showpacs,preprintnumbers,
 amsmath,amssymb,
 aps,
pre,
]{revtex4-1}

\usepackage{graphicx}
\usepackage{dcolumn}
\usepackage{bm}
\usepackage{hyperref}
\begin{document}

\title{States of motion in an AC liquid film motor: experiments and theory}

\author{A. Amjadi}
\affiliation{Physics Department, Sharif University of Technology, Tehran, Iran.}
\author{R. Nazifi}
\affiliation{Physics Department, Sharif University of Technology, Tehran, Iran.}
\author{R. M. Namin}
\affiliation{Department of Mechanical Engineering, Sharif University of Technology, Tehran, Iran.}%
\author{M. Mokhtarzadeh}
\affiliation{Physics Department, Sharif University of Technology, Tehran, Iran.}

\date{\today}

\begin{abstract}

The liquid film motor (LFM) is a simple device which serves as a laboratory to conduct experimental
research in basic theoretical studies of electrohydrodynamics (EHD). In addition the LFM can play an
important role in technological applications such as micro mixers and washers.
In this paper we initially performed experiments to examine the theory regarding rotation and vibration
of an AC LFM. While many theoretical predictions are in agreement with our experimental data, but in
the threshold of vibration there is an obvious disagreement; we showed that no threshold exists and
oscillation is observed in any field value. Experimental evidence showed the existence of an elastic
deformation in water before reaching the yield stress and lead us to the use of the elastic Bingham
fluid model. We revised the theories using this model and investigated the behaviour of a vibrating
LFM which has more applications in mixing. We applied a numerical solution to derive different phases
of vibration and distinguish between elastic and plastic vibrations, the latter which could be used
for mixing.

\begin{description}
\item[PACS numbers]
47.65.-d, 68.15.+e, 83.60.La
\end{description}

\end{abstract}

\maketitle
\section{Introduction}

Quasi two-dimensional suspended liquid films serve as a laboratory to study hydrodynamical characteristics of two dimensional fluid flows, which has received intensive attention \cite{chomaz1990soap,couder1989hydrodynamics,rivera2000external,PhysRevLett.86.4326,langevin2011surface,sonin1998freely,sheludko1967thin}.
Applying an electric field to liquid or liquid crystalline films can produce electro-hydrodynamical(EHD) flows in the films, a topic which has attracted much interest due to its wide 
range of applications specially in micro and nano scale \cite{espin2013electrohydrodynamic, shirsavar2012rotational, zhao2012advances,sun2011ac,lee2011microfluidic,dittrich2006lab,stone2004engineering,morris1990electroconvection}.
The applications involve microfluidic devices such as dispensers, mixers, motors, separators, pumps and other functional units that are needed for a lab on a chip systems \cite{stone2004engineering,dittrich2006lab} or drug delivery, environmental and food monitoring, biomedical diagnoses and chemical analyses, etc \cite{zhao2012advances}.
Our goal in this paper is to study the physics of such EHD phenomenon in the case of the liquid film motor (LFM).

\begin{figure}
\includegraphics[width=0.8\linewidth]{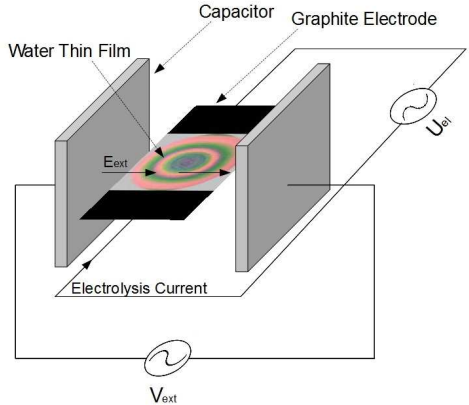}
\caption{General Experiment Setup}
\label{fig:setup}
\end{figure}

We previously introduced an experimental device which can induce pure rotations in water suspended films \cite{amjadi2009, shirsavar2011electrically, shirsavar2011electrically, shirsavar2012rotational}.
The device is consisted of a two-dimensional electrolysis frame connected to a pair of electrodes which conduct electric current through the liquid film. The frame is located inside a large parallel plate capacitor which produces external electric field in the plane of the film (figure \ref{fig:setup}). By changing the magnitudes of the electric current and the applied external electric field, the threshold conditions for the onset of rotation has been obtained. The direction and the speed of rotation can be controlled through manipulation of the direction and the strength of the current and/or external electric field. Our experiments showed that the rotation direction of the induced vortices is strictly related to the directions of the applied fields and it obeys the right-hand rule ($\vec{E}_{ext}\times \vec{J}_{el} $). We have observed that the rotation exists for polar liquids like Aniline, Anisole and Chlorobenzene. The experiments show that the threshold of the fields for the on-set of rotation is in the same order of magnitude for many polar liquids \cite{shirsavar2011electrically}, irrelevant to their conductivity, viscosity and density values. We have also studied the effect of phase and frequency on the rotation of the film \cite{shirsavar2011electrically}. On AC LFM, when external electric field $E$, and electrolysis current $J$, are in phase having the same frequencies, the direction of the fields reverse alternatively and simultaneously so that the direction of rotation remains unchanged (direction of $ \vec{E}_{ext}\times \vec{J}_{el} $ remains unchanged). If the fields would have different frequencies, the application of external and internal fields to the film cannot produce rotating vortices but it causes vibrational movements. If the applied alternating fields have exactly the same frequencies the liquid film rotates. In this case, the threshold and velocity of the rotation depend on the time phase difference between the fields and their magnitudes.

The remote and reversible control over the fluid can be used in mixing of confined liquids in miniaturized devices. Our efficient control over the fluid can be used in active mixers which use external forces for mixing the samples in special microchannel devices \cite{lee2011microfluidic}. The mixing can be achieved via the application of noncontact electrodes to microfluidic films. The speed and direction of the mixing can be controlled via the strength and direction of the electric fields \citep{amjadi2009} and it can be used to regulate the mixing speed.

The explanation of the rotation mechanism is important in understanding the physics behind the phenomenon.
Many attempts have been made to explain the phenomenon in terms of electrohydrodynamics.
In 2009, \citet{shiryaeva2009theory} suggested a detailed mathematical model on the phenomenon. Their physical
perspective was based on the fact that the external filed induces electrically charged areas at the boundaries
of the film, and the electric current exerts force and causes rotation. Their theory
is based on the edge effects, and the rotation predicted therein is maximum at the edges which is not
consistent with experiments. In 2011 \citet{liu2011dynamical} suggested another perspective on this phenomenon.
They suggest that the rotation is caused by the continuous competition between the destruction of
the polarization equilibrium by the electric current and it's reestablishment by the external electric field.
Their explanation is supported by the picture water in quantum field theory, the existance of
coherent domains in which all water molecules oscillate in phase between two configurations. They also
explained the fact that a minimum field is needed for the rotation by assuming water film to be a
Bingham plastic in presence of an electric field.

The quantum field theory has also been used in explaining other electrohydrodynamical phenomena such as the
floating water bridge \cite{del2010collective}, which is the suspension of a water thread between two
beakers in a high voltage regime \cite{fuchs2007floating}. But in the case of the floating water bridge, much
more attention has been given to theoretical explanations based on classical electrohydrodynamics,
neglecting the need of a quantum field theory explanation for that phenomenon 
\cite{widom2009theory, saija2010communication, marin2010building, aerov2011water, morawetz2012theory},
and are shown to be in agreement with experiments \cite{namin2012equilibrium}.

In a more recent attempt, \citet{liu2012water}, presented a model for dynamical characteristics of the AC water film motor. They derived a series of the specific features of this motor, e.g. the threshold conditions for the on-set of vibration and rotation of the film and the influence of the frequencies of the AC fields and their phase differences on EHD motions. Their results for the threshold conditions of the AC fields for the onset of rotation are in agreement with our previous experiments \cite{amjadi2009}.Their model  predicts some features which have to be experimentally verified, 
such as three states of motion in case the fields are in a same frequency but a different phase: rest, vibration and rotation. In case
the fields have different frequencies, the theory predicts two states: rest and vibration with specified frequencies.

In this paper, we first construct precise quantitative experiments to test the
theoretical predictions in section \ref{sec:exp}. We make use of an image processing
technique for the velocimetry, and detect the vibration using Fourier transformation.
The experimental results in sec. \ref{sec:exp} mismatch with the theoretical results,
therefore we find the mismatch source and develop a revised theory in section \ref{sec:theory}
and derive the results by using a numerical solution, leading to a conclusion in section \ref{sec:conc}

\section{Experiments}
\label{sec:exp}

\begin{figure}
\centering
\includegraphics[width=\linewidth]{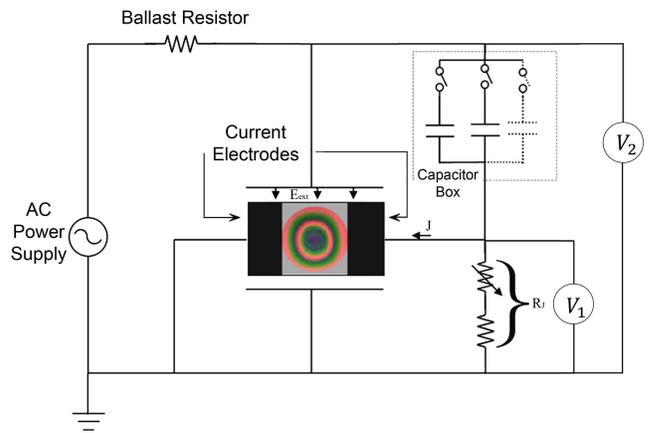}
\caption{Experiment setup for fields with same frequencies and a phase difference }
\label{fig:setupa}
\end{figure}

We construct quantitative experiments to examine the theoretical predictions in an AC LFM.
The theory \cite{liu2012water} explaines the dynamical behaviour of the film when AC voltages are applied, and in the cases where
the applied electric fields have different frequencies or time phases, it predicts the existence of various states of motion
in different voltage products. 

The general setup of the device consists of a 2D frame with two electrodes on the sides that conduct electric current  $J_{curr}$ through the liquid film. The electrodes are subjected to an AC voltage. The liquid film is formed between these electrodes in a circular shape with the diameter of 8.1 mm.
In order to apply an AC external electric field to the film, the frame is located between two large parallel metal plates acting like a capacitor (Fig. \ref{fig:setup}).
In all of our experiments the direction of the electric current $E_{curr}$   is perpendicular to the direction of the external electric field  $E_{ext} $ to maximize the strength of vibration.

In the first part of our experiments, the electrodes conducting current $J_{curr}$ and the plates producing external electric field $E_{ext}$ are subjected to the same power supply, in order to have fields with equal frequencies of 41 Hz. The phase differences between the fields are generated by a simple Series-Resistance-Capacitance Circuit, as follows:
The voltage across the resistance $V_R$ is connected to the electrodes which produce the electric current, $J_{curr}$ while the AC power supply is directly connected to the capacitor to produce $E_{ext}$ (figure \ref{fig:setupa}). By changing the capacitance value of the capacitor, we can produce different phase angles between the electric current and the external electric field. We have tested phase angles of 60, 75 and 85 degrees.

\begin{figure*}
\includegraphics[height=4cm]{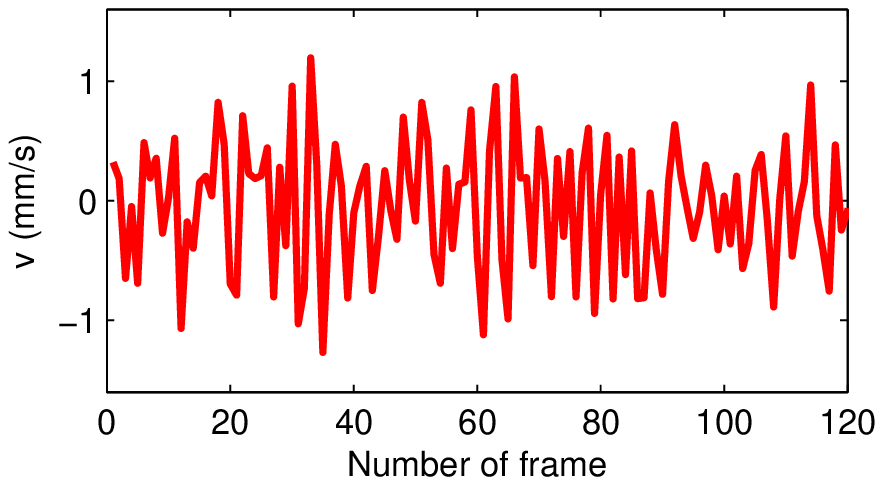}
\includegraphics[height=4cm]{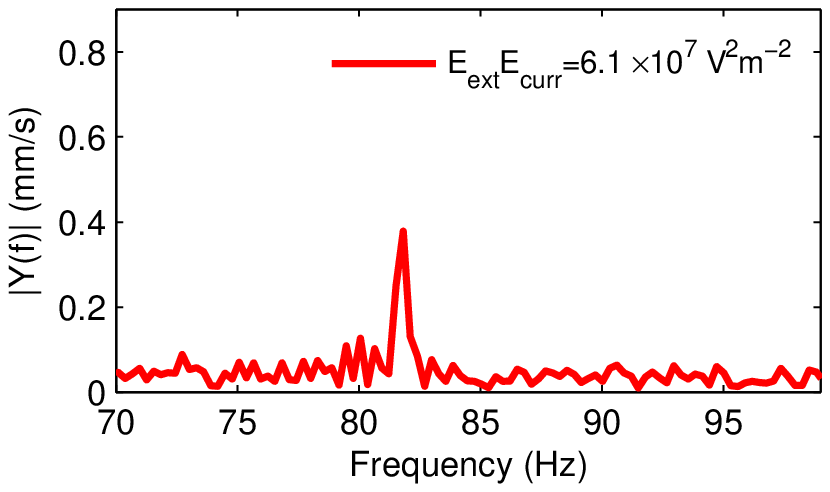}\\
\includegraphics[height=4cm]{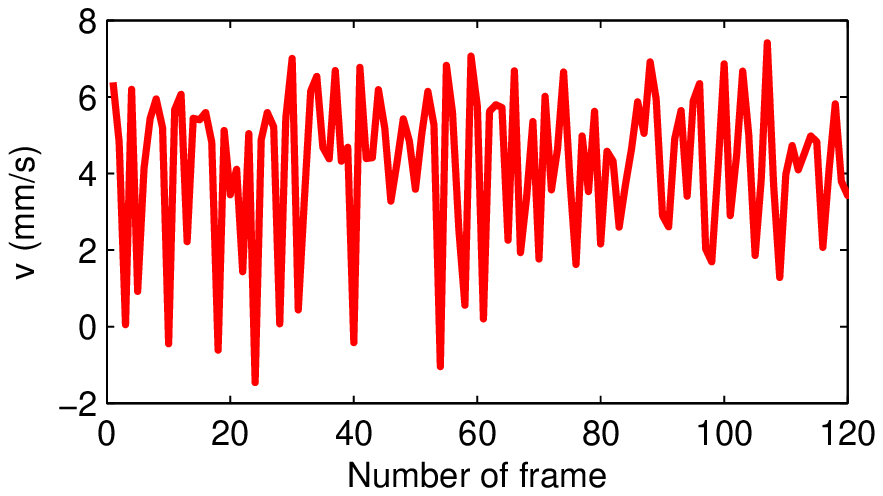}
\includegraphics[height=4cm]{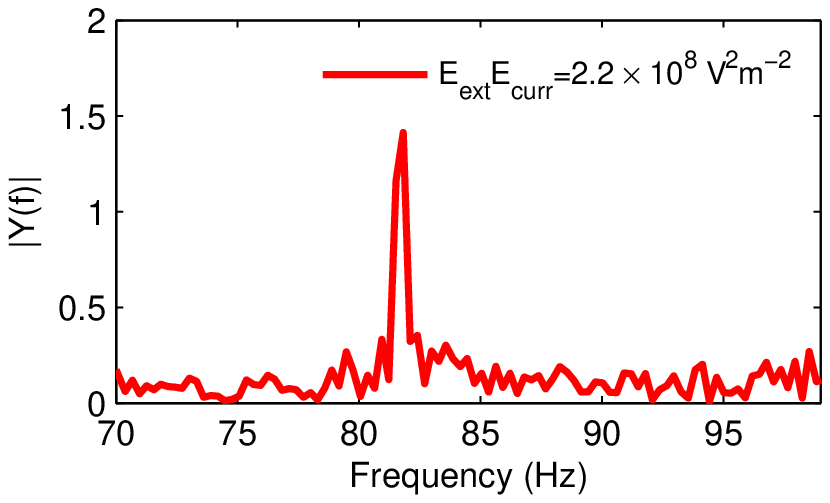}
\caption{Experimental results for the case of electric fields with a phase difference $\phi = 75^\circ$ in a frequency of 41Hz.
Left: average tangential velocities.
right: amplitude of vibration in the frequency domain obtained by Fourier transformation.
The electric fields in the top plots are below the predicted threshold of rotation and only vibration is visible.
The fields in the bottom blots are higher than the threshold of rotation and an average rotation is detected.}
\label{fig:exp_res1}
\end{figure*}

The vibration is quite hard to detect in experiments. The vibration displacements and velocities are not high
enough to be detected with naked eye, especially near the expected thresholds of vibration.
For this reason a velocimetry program was developed. A method similar to the Particle Imaging Velocimetry
(PIV) was applied, with the difference that no tracer particle was used. The color patterns traced on the film were fed into the pattern matching algorithm, and the difference between the patterns of two
subsequent frames was used to find the velocity vector field between each two frames.  We used a modified
version of the open source MatPIV 1.6.1 \cite{matpiv} which calculates the velocity vector field 
by a window-shifting technique for the pattern matching.  The tangential
velocities were calculated at each point and averaged to find $v(t)$, as plotted against time in figure \ref{fig:exp_res1}.
The vibration was detected by applying a fast Fourier transform to $v(t)$ in a period of 10 seconds.

Our experiments verified the existence of a rotatory state of motion in electric fields higher than a specific threshold. We observed that
in this case, the rotation is along with a slight vibration as shown in figure \ref{fig:exp_res1}. In the case of a smaller electric field products than
the threshold of rotation, the motion becomes pure vibration as predicted in the theory \cite{liu2012water} and shown in figure \ref{fig:exp_res1}.
The theory predicts that in smaller electric fields, the motion changes to absolute rest, but our experiments did not enable us to examine
this fact because the displacements and velocities become smaller than our measuring accuracy.

\begin{figure*}
\includegraphics[height=4cm]{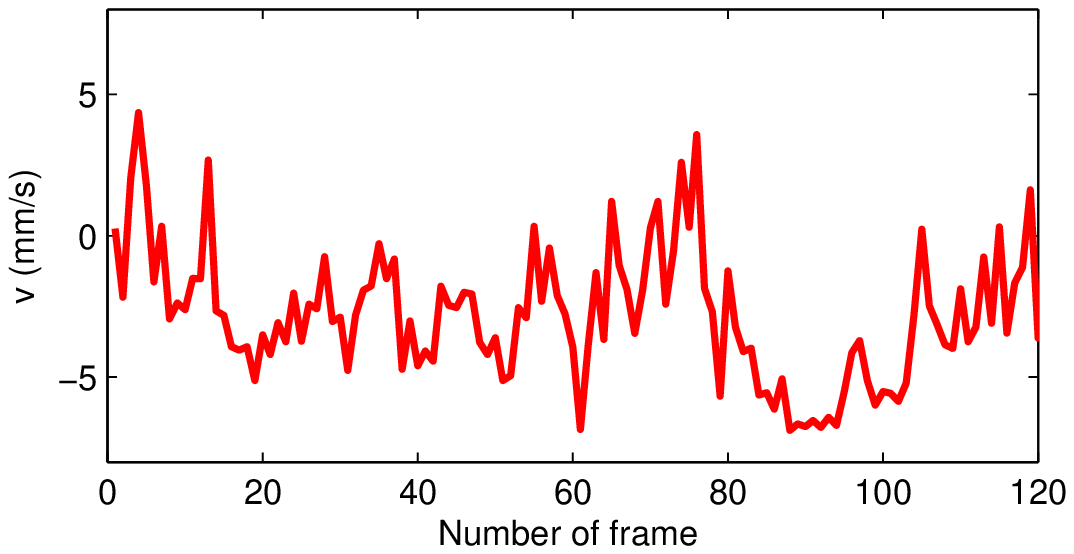}
\includegraphics[height=4cm]{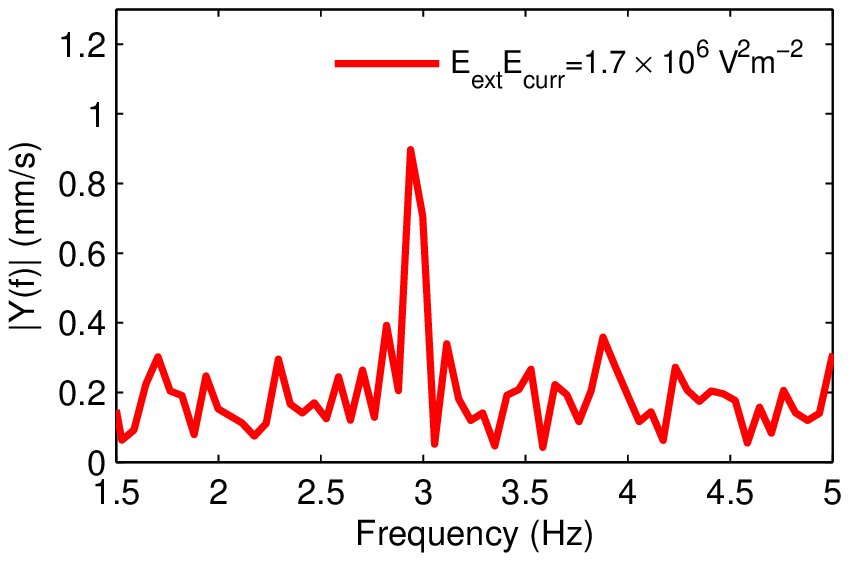}\\
\includegraphics[height=4cm]{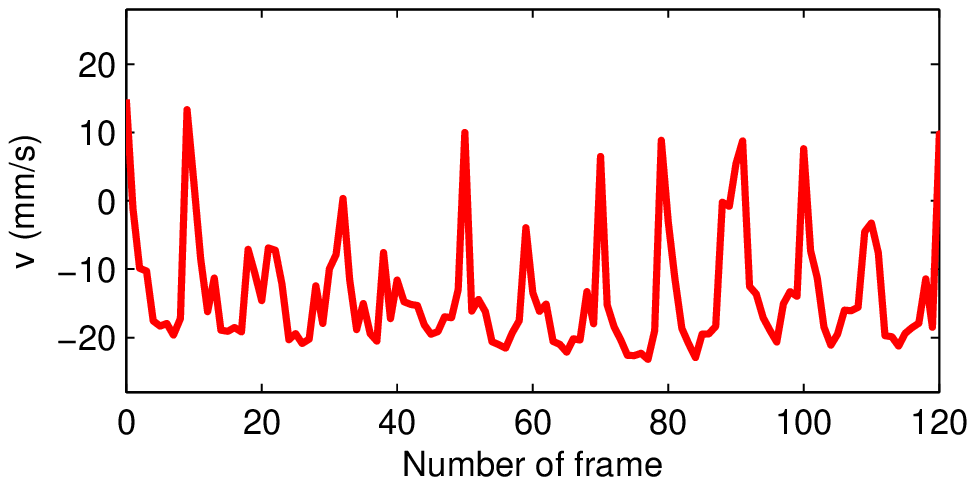}
\includegraphics[height=4cm]{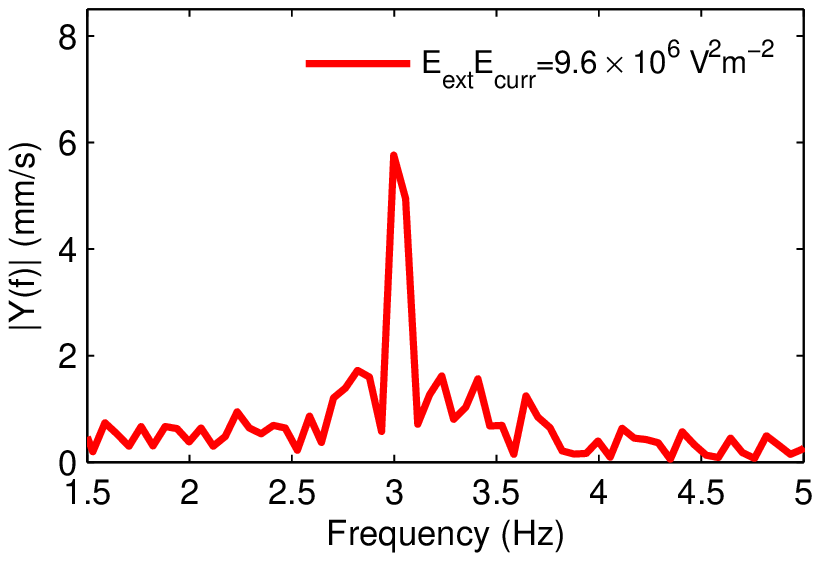}\\
\includegraphics[height=4cm]{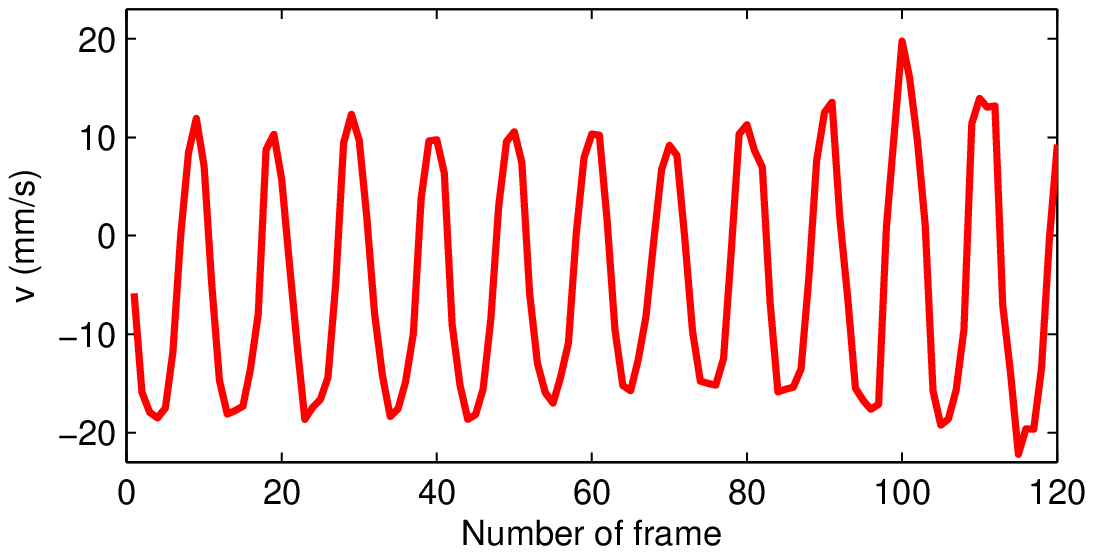}
\includegraphics[height=4cm]{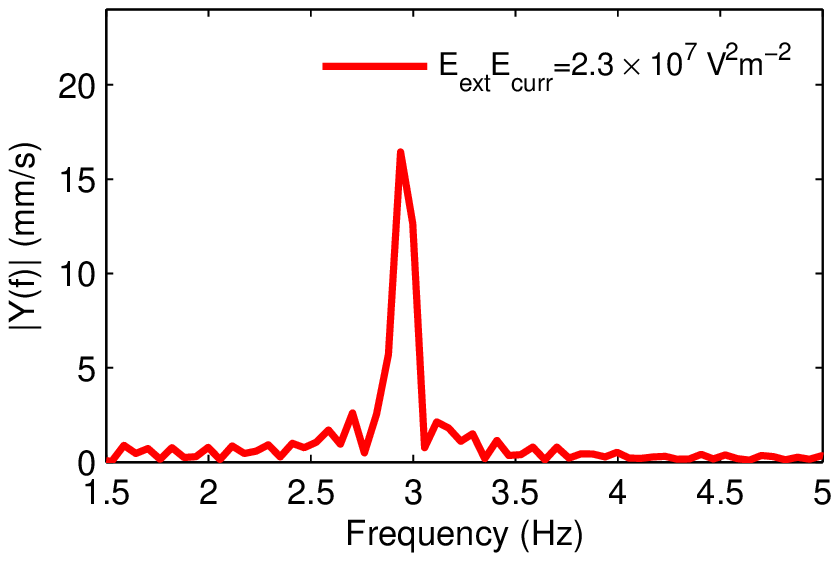}\\
\includegraphics[height=4cm]{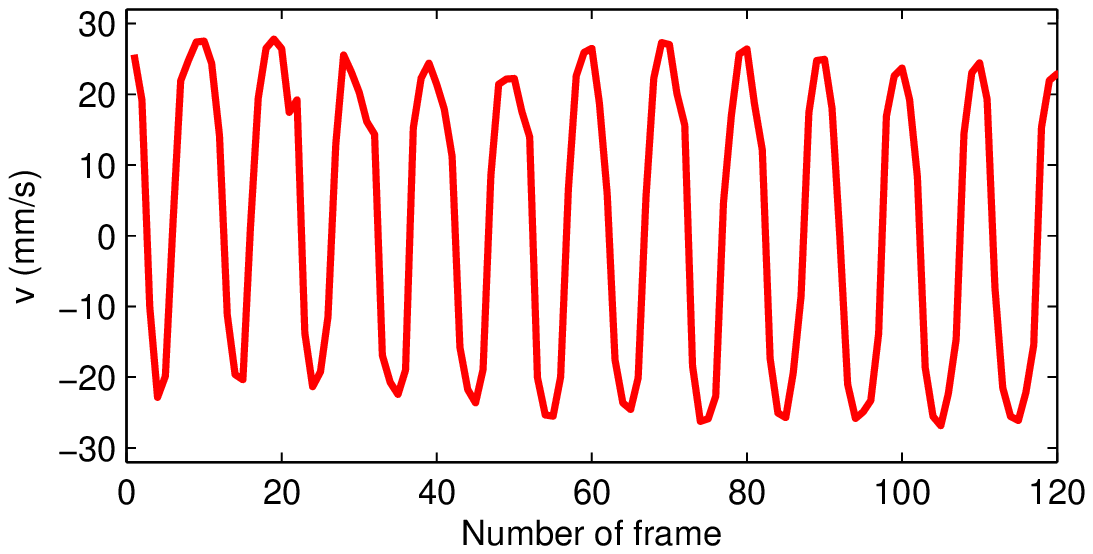}
\includegraphics[height=4cm]{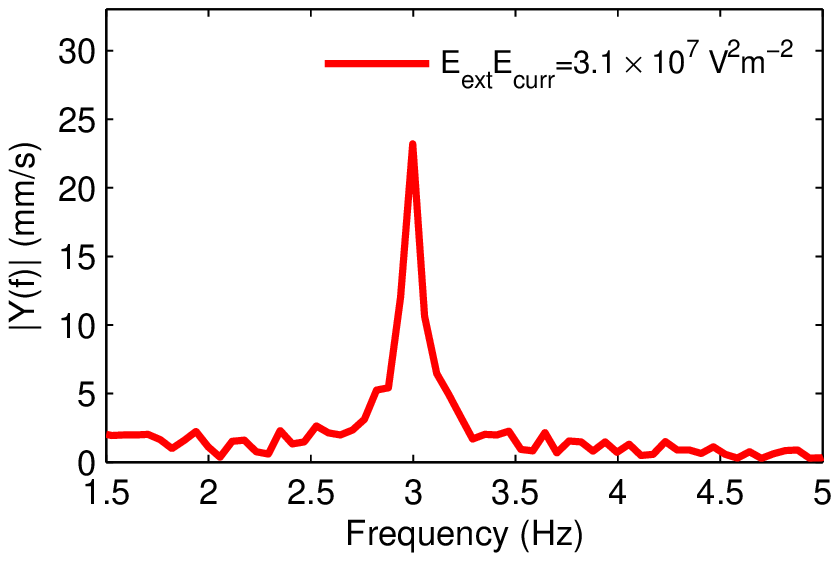}
\caption{Experimental results in case of electric fields with different frequencies of 50Hz and 53Hz.
Left: average tangential velocities vs. time.
right: right: amplitude of vibration in different frequencies obtained by Fourier transformation.
The ratio between the product of of electric fields to the theoretical threshold of vibration according
to \cite{liu2012water} is 0.1, 0.6, 1.2 and 1.6 from top respectively. The plots show clearly
that the threshold of vibration is not correctly defined, as the two top plots are far below the predicted vibration threshold.}
\label{fig:exp_res2}
\end{figure*}

In the second set of experiments, different frequencies were applied to the electrodes and the capacitor. We used the frequency of
53Hz to the electrode and the frequency of 50Hz to the capacitor, thus expecting a vibration of 103Hz and a vibration of 3Hz
according to the theory and in the FFT of $v$ this frequency was clearly detected. The theory predicts that for voltages smaller than
a specific threshold the Bingham plastic causes a state of rest, where no rotation or vibration occurs.
Our experiment results showed that
this prediction of the theory is not correct, i.e.  the state of rest does not exist and vibration was clearly detected in the specified
frequency in electric fields far below the thresholds predicted theoretically, as shown in figure \ref{fig:exp_res2}. Thus the first stage of motion in
the AC liquid film motor predicted theoretically \cite{liu2012water} does not exist.

Our other experiments in the DC motor show that when the product of applied voltages is less than the threshold
of rotation, a slight angular displacement is observed when the voltages are applied, and a reverse
angular displacement is observed when the voltages are turned off. This experimental evidence shows
that an elastic deformation takes place when the stresses applied are less than the yield stress i.e.
in the case that the Bingham plastic does not show a plastic behaviour, it is not rigid, but deforms
elastically when a stress is applied. Based on this explanation, we revisit the theoretical
predictions for the AC liquid film motor and study the vibration behaviour in section \ref{sec:theory}.

\section{Theory revision}
\label{sec:theory}

Considering the liquid film to be a Bingham plastic in presence of the external electric field
(as \cite{liu2011dynamical, liu2012water})
explains well the steady phenomenon of rotation in a LFM and the thresholds of rotation.
This assumption is explained using the quantum field theory; the plasticity of the film is
suggested to be caused by long-ordered chains composed of coherent domains with an electric
dipole moment resulting from the alignment of the electric dipole moments of their coherently
oscillating molecules \cite{liu2011dynamical}.
In the case
when the shear stress is more than a minimum shearing stress ($\tau < \tau_0$) the fluid shows 
no displacement.

However this assumption does not correctly model the unsteady case of an AC LFM as shown
experimentally in section \ref{sec:exp}. In this case, the fluid motion can not be classified
in two cases of 'no shear rate' and 'viscous shear'. Considering the film to be an elastic Bingham
fluid is a more accurate assumption for the unsteady case and describes well the experimental data
on rotation and vibration of LFM. In this case when
the layers of the fluid do not exhibit a plastic flowing deformation, an elastic solid behaviour is 
expected, i.e. an elastic reversible deformation occurs when a shear stress is applied. It has been 
shown by \citet{yoshimura1987response} that
elastic Bingham fluids exhibit a vibration when subjected to a variable stress.

To formulate this effect quantitatively, we define an elastic shear strain as a function
of time and radius, $\gamma(r, t)$. A shear stress $\tau_s$ is applied between each layer which follows the linear
relation $\tau_s = G \gamma$ in which $G$ is the shear modulus of water in the elastic range.
$\gamma$ has a limited range of $\vert \gamma \vert < \gamma_{max}$ and if the deformation exceeds 
this range, the residual deformation is a plastic deformation. $\gamma_{max}$ is the strain where 
the maximum elastics stress ($\tau_0$) is applied, thus $\gamma_{max} = \tau_0 / G$.

To define the motion of the film from this point of view, the parameters would be the tangential
velocity $u(r, t)$ and the elastic shear strain $\gamma(r, t)$. The torque exerted to a disc liquid
film with a radius of $r$ and thickness of $h$ is consisted of three moments:
\begin{eqnarray}
M_{rd} = M_{curr} + M_{Bd} + M_e
\label{eq:moment}.
\end{eqnarray}
where $M_{curr}$ is the moment caused by the dipole moments and the steady electric field of the
electric current, $M_{Bd}$ is caused by the viscosity force and $M_e$ is because of the elastic
deformation. According to \citet{liu2011dynamical}, $M_{curr}$ follows this relation for our
system:
\begin{eqnarray}
M_{curr} = M_{DE} \pi r^2 h
\label{eq:m_cur},
\end{eqnarray}
where $M_{DE}$ is the torque per unit volume and is equal to 
$\varepsilon_0 (1 - 1 / \varepsilon_r) E_{ext} E_{cur}$ in which
$\varepsilon_0$ is vacuum permittivity, $\varepsilon_r$ is the relative permittivity of the
liquid, $E_{ext}$ is the external field applied by the capacitor, $E_{curr}$ is the field
causing the electrical current normal to the external field and $h$ is the thickness of the film.

The viscous and elastic moments can be calculated using the shear stresses of each effect:
\begin{eqnarray}
M_{Bd} = 2 \pi r^2 h \mu (\frac{\partial u}{\partial r} - \frac{u}{r})
\label{eq:m_bd},
\end{eqnarray}
\begin{eqnarray}
M_e = 2 \pi r ^ 2 h  G \gamma
\label{eq:m_e}.
\end{eqnarray}

The moment exerted to a ring element of the film with radius $r$ to $r + dr$ can be calculated
by the derivation of the total torque $M_{rd}$, which results in:
\begin{eqnarray}
dM = dV [\mu ( r \frac{\partial^2 u}{\partial r^2} + \frac{\partial u}{\partial r} - \frac{u}{r}) \nonumber \\
	+ G (r \frac{\partial \gamma}{\partial r} + 2 \gamma)
	+ M_{DE}]
\label{eq:dm}.
\end{eqnarray}
where $dV$ is the volume of the ring element.
This moment will cause an angular acceleration, which can be written in terms of the derivation
of u, thus leads to:
\begin{eqnarray}
\dfrac{1}{r}\dfrac{\partial u}{\partial t} = \dfrac{dM}{\rho r^2 dV}
\label{eq:rot}.
\end{eqnarray}
where $\rho$ is the density of the fluid.

For the shear strain there must be an equation to be coupled with equation \ref{eq:rot} and form the
differential equation of motion. The shear strain rate (S) can be calculated as the difference 
between the rotation of a specific point and the overall rotation of the segment:
\begin{eqnarray}
S(r, t) = \frac {\partial u} {\partial r} - \frac{u}{r}
\label{eq:strain}.
\end{eqnarray}
but this shear strain is not totally the elastic strain: as it is explained before, the elastic
strain must be in the limited range. So the residual strain will be a plastic strain. Thus the
derivation of the elastic strain can be formulated as follows:
\begin{eqnarray}
\frac {\partial \gamma} {\partial t} = \left\{\begin{matrix} 0 & 
 , &\gamma = \gamma_{max} & \land & S > 0 \\ 
0 & , &\gamma = -\gamma_{max} & \land & S < 0 \\
S & , &-\gamma_0 < \gamma < \gamma_0 &  \lor & \gamma S < 0
\end{matrix}\right.
\label{eq:gamma}
\end{eqnarray}

Equations \ref{eq:gamma} and \ref{eq:rot} form the equations of motion for this system.
The boundary conditions for this system is that flow velocity is zero because of the
no-slip condition, and there are no shear velocity and shear deformation at the center because of
symmetry:
\begin{eqnarray}
\begin{matrix} 
u(t, R) = 0 & u(t, 0) = 0 \\
\gamma(t, 0) = 0 & \frac {\partial \gamma} {\partial r}(t, R) = 0 \\ 
\end{matrix}
\label{eq:boundary}
\end{eqnarray}
also the initial conditions are:
\begin{eqnarray}
\begin{matrix} 
u(0, r) = 0 & \gamma(0, r) = 0
\end{matrix}
\label{eq:initial_c}
\end{eqnarray}

\begin{figure}
\includegraphics[width=\linewidth]{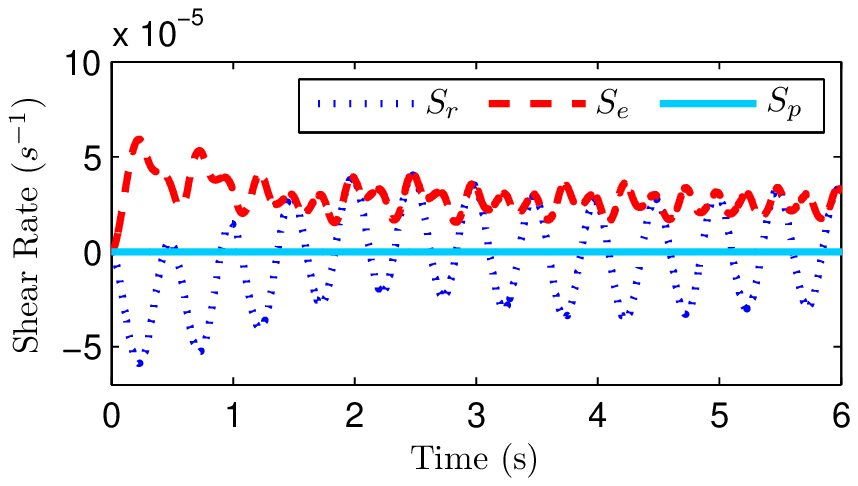}
\includegraphics[width=\linewidth]{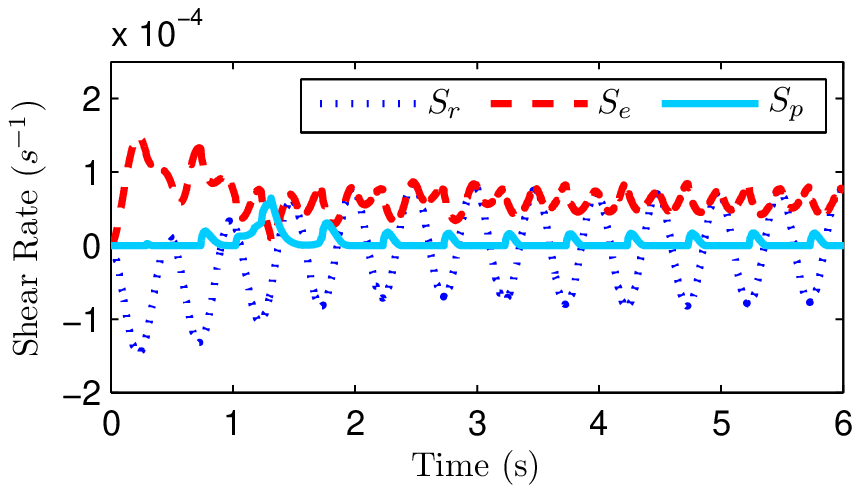}
\includegraphics[width=\linewidth]{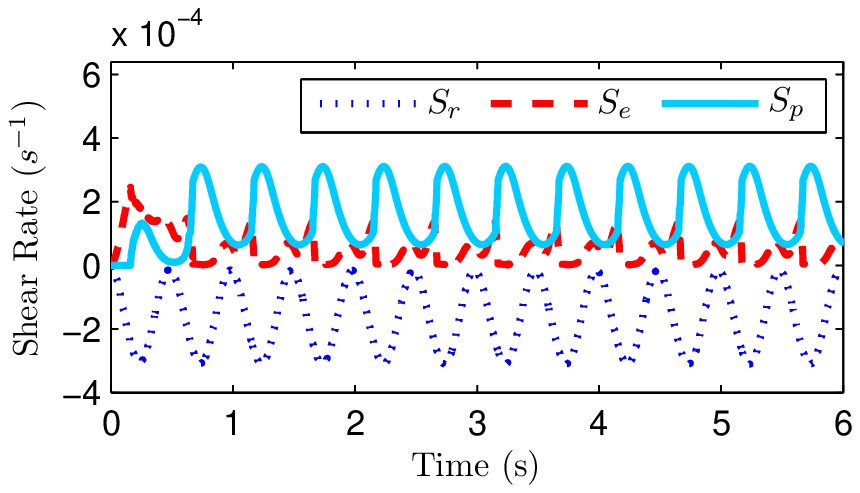}
\caption{Numerical results of shear rates as a function of time for the case of phase difference between the electric fields: $\phi = 60^\circ$ and the frequency of 1Hz.
Three regimes can be seen:
top: Elastic vibration: ${E_0}_{curr} {E_0}_{ext} = 2 \times 10^{7} V^2 m^{-2}$,
middle: Plastic vibration: ${E_0}_{curr} {E_0}_{ext} = 5 \times 10^{7} V^2 m^{-2}$,
bottom: Rotation: ${E_0}_{curr} {E_0}_{ext} = 10 \times 10^{7} V^2 m^{-2}$.}
\label{fig:theory1}
\end{figure}

\begin{figure}
\includegraphics[width=\linewidth]{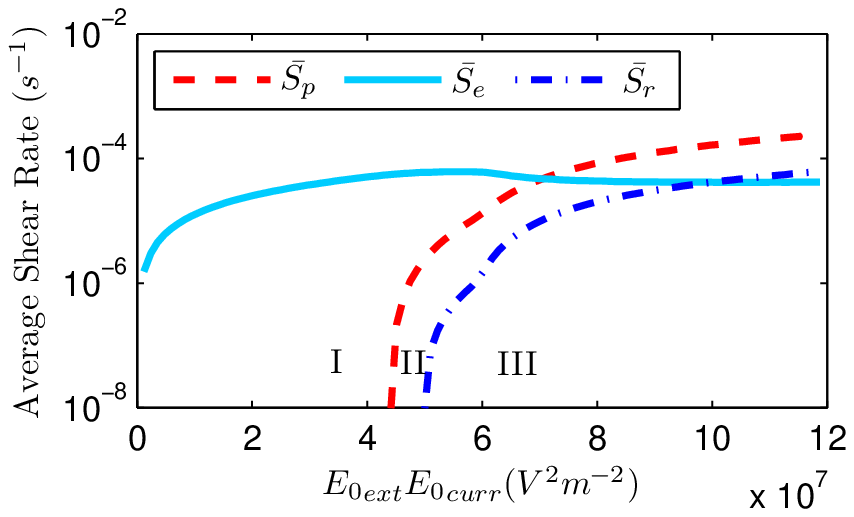}\\
\includegraphics[width=\linewidth]{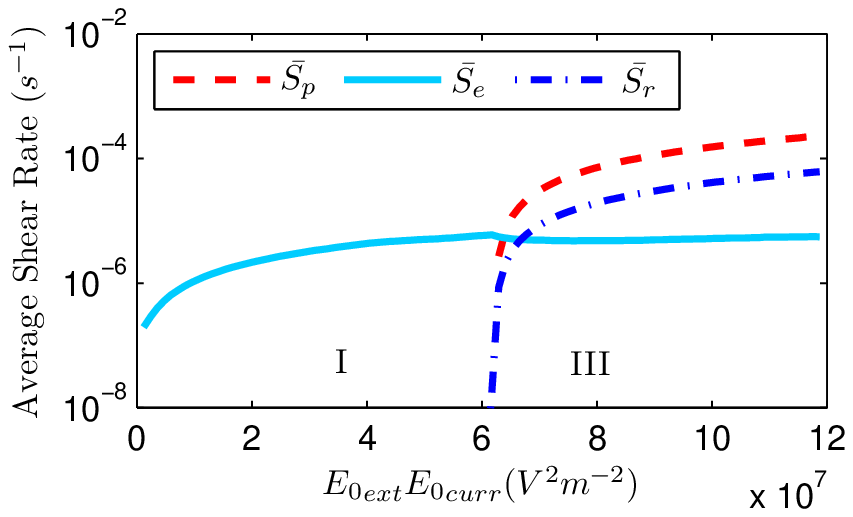}
\caption{Numerical results for average shearing rates as a function of the product of the applied electric fields. Phase difference = 60 degrees.
top: Frequency = 1 Hz. The three states of motion can be seen: I) Elastic vibration, II) Plastic vibration III) Rotation.
bottom: Frequency = 50 Hz. The state of plastic vibration diminishes in high frequencies.
\label{fig:theory2}}
\end{figure}

To solve the equations
and obtain results, we applied a numerical
method solving the ordinary differential equation system using an explicit Runge-Kutta (4,5) formula,
the Dormand-Prince pair \cite{dormand1980family}.

The results are highly dependant on the amount of maximum shear strain assumed. In experiments the elastic
displacement of the film in connecting and disconnecting the electric fields was clearly visible, so we
have estimated the maximum shear angle to be in the order of one degrees.

For the case where the applied voltages are of a similar frequency and have a phase difference, our simulation
shows the existence of two different possible stages of motion: vibration with no net rotation and rotation
plus vibration. Note that as explained above, the state of rest does not exist. In addition to this,
the state of vibration can be divided into two different cases: one where the vibration exhibits a plastic motion,
or where it is caused only by elastic deformation. The difference is not distinguishable in experiments, but it has
high impact in mixing application, i.e. a vibration with elastic deformations is not a cause of 
detachment of the fluid particles and is not suitable for mixing. To evaluate
this effect quantitatively, we define the elastic shear rate (${S_e}$) and plastic shear
rate (${S_p}$) in the liquid film as follows respectively:
\begin{eqnarray}
S_e = \frac{1}{R} \int_0^R |\frac {\partial \gamma} {\partial t}|  \mathrm{d}r,\\
\label{s_e}
S_p = \frac{1}{R} \int_0^R (|S -\frac {\partial \gamma} {\partial t}|) \mathrm{d}r,
\label{s_p}
\end{eqnarray}
and also define a rotatory shear rate
that is:
\begin{eqnarray}
S_r = \frac{1}{R} \int_0^R S \mathrm{d}r
\label{s_r}.
\end{eqnarray}
The averages of the shear rates in a full period in a time when the motion has reached a steady rotation and
vibration is defined as $\bar{S_e}$, $\bar{S_p}$ and $\bar{S_r}$ respectively.
Monitoring these values in specific fields, phase differences and frequencies leads to an estimation on the
amount of elastic vibration, plastic vibration and rotation in a state of motion.
Figure \ref{fig:theory1} shows this behaviour in three cases. In the case that the frequencies are the same
but with a phase difference, there might be three states of motion: elastic vibration, plastic vibration and
rotation. Note that the state of rest predicted by \citet{liu2012water} has been replaced by the state of
elastic vibration. The threshold between elastic and plastic vibrations depend on the frequency of the
fields, i. e. in a high frequency, there will not be enough time to deform and reach the maximum strain to
lead to a plastic deformation even though the stress is large enough to cause a plastic deformation. This fact has
been visualized in figure \ref{fig:theory2}, comparing $\bar{S_p}$, $\bar{S_e}$ and $\bar{S_r}$ for two cases of $50Hz$ and
$1Hz$. It is also seen that in high frequencies, the state of plastic vibration gets rare and the motion
shifts from elastic vibration to rotation right away by increasing the fields. The plastic vibration is of special importance
for it can be used for micro mixing \cite{liu2013water}, and our results show that it is rare in high
frequencies. 

\begin{figure}
\includegraphics[width=\linewidth]{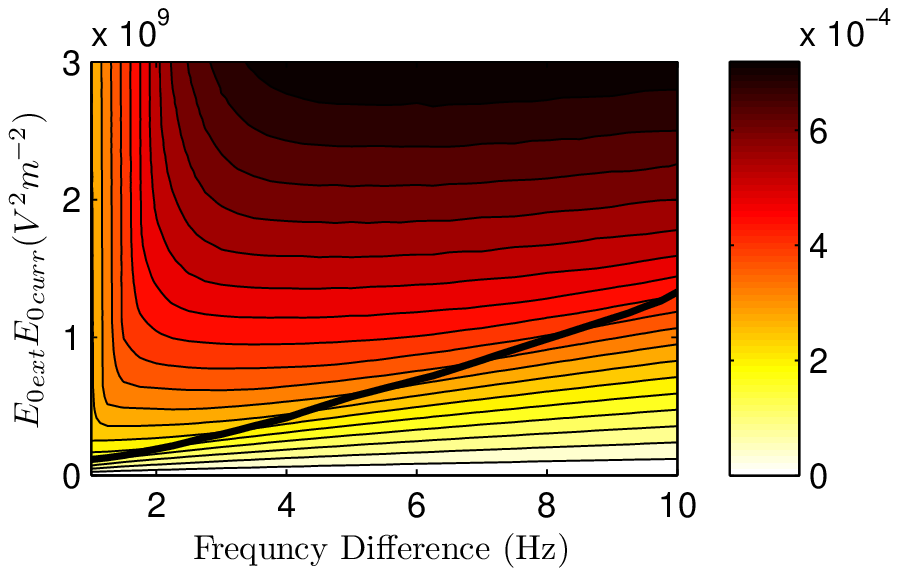}
\includegraphics[width=\linewidth]{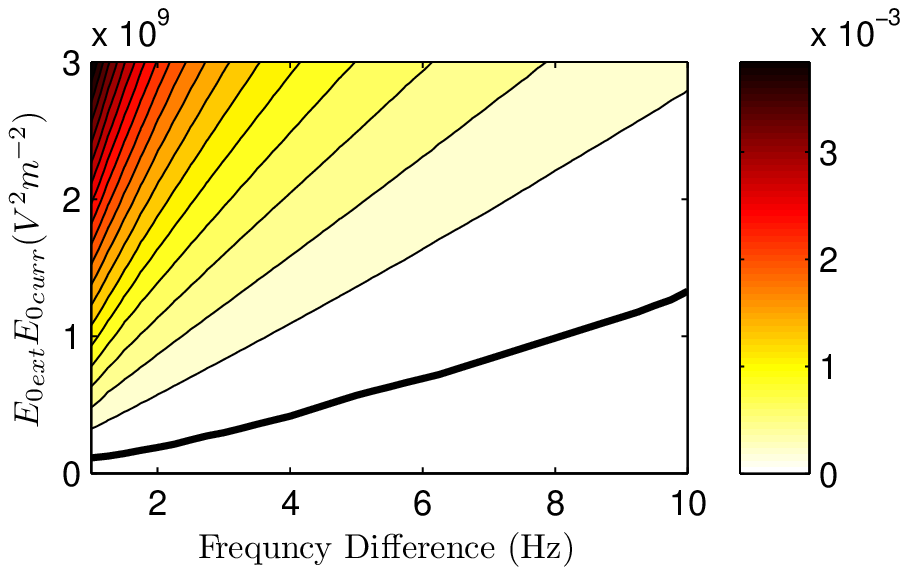}
\caption{top: Contour of average elastic shearing rate ($s^{-1}$), bottom: Contour of average plastic shearing rate ($s^{-1}$). The black solid line shows threshold of plastic vibration, which only above that is suitable for mixing .}
\label{fig:theory3}
\end{figure}

Our theory also leads to changes of results in the case when the frequencies of the 
two applied fields are different. The state of rest will disappear, and the vibration
will be consisted of two phases of elastic and plastic vibration. A threshold will
appear that with electric fields smaller than that, no plastic vibration will be present.
As shown in figure \ref{fig:theory3}, this threshold increases in higher frequencies, i.e. with the
increase of frequency, a higher electric field is needed to cause a plastic deformation
because deformations and velocity changes occur in a smaller time. Figure \ref{fig:theory3} shows how
the amount of elastic and plastic vibrations change in different frequencies and electric
fields.

\section{Conclusion}
\label{sec:conc}

We performed experiments to examine the predictions of theories regarding the AC liquid film motor. We
constructed the setup for the AC fields with different frequencies and used
a velocimetry image processing to obtain the vibration velocity.
Many predictions of the theories of Liu et al. \cite{liu2011dynamical, liu2012water} are the same as our experimental results,
in the case of same frequencies with phase difference between the electric fields and in the case of frequency differences the rotation and vibration were observed and the vibration frequency
matches the predictions therein. It was observed that in the case of phase differences, vibration
and rotation exist simultaneously and by increasing the electric fields, the ratio between vibration
and rotation velocities change. In high electric fields, the rotation will be dominant while in
lower fields the vibration dominates.

Our experimental results also showed that in the case of different frequencies the state
of rest which was predicted in the theory did not occur and vibrations exist in any electric field
strengths.
We found that for explaining this inconsistency it is adequate to apply an elastic Bingham fluid model instead of the previous
Bingham plastic model in the theory.
We revised the theory considering this model and used a numerical solution. The results are in good agreement with our experimental findings.

We define a maximum elastic shearing strain for water in electric fields and distinguish between the elastic and plastic vibrations. The plastic vibration
can be used for application in mixing, while the elastic vibration is not.
Our theoretical results show that in high frequencies for the case of a phase difference, by increasing the fields the state of elastic
vibration changes to rotation and the state of plastic vibration diminishes.

As a final conclusion in the application of mixing, lower frequencies and higher electric field strengths are required,
otherwise an elastic vibration may occur with no benefit on mixing. The threshold of the frequency  and electric fields
depend on the dimensions and the electric properties of the liquid.

\begin{acknowledgments}
We greatly thank Dr. R. Shirsavar and Prof. S. W. Morris for their helpful comments at the initial
stages of the work and also
Dr. Z-Q. Liu for the discussions about this effect and his helpful remarks.
This work is funded by Applied Physics Research Center of Sharif University of technology and all the 
experiments were performed at the
Medical Physics Laboratory of the Physics Department.
\end{acknowledgments}

\bibliography{LFM4}

\end{document}